# PATTERNS FOR BUSINESS-TO-CONSUMER E-COMMERCE APPLICATIONS

Xiaohong Yuan[1] and Eduardo B. Fernandez[2]

[1]Department of Computer Science, North Carolina A&T State University
Greensboro, NC, USA
xhyuan@ncat.edu
[2]Department of Computer Science and Engineering, Florida Atlantic University
Boca Raton, FL, USA
ed@cse.fau.edu

### ABSTRACT

*E-commerce is one of the most important web applications. We present here a set of patterns that describe shopping carts, products, catalogue, customer accounts, shipping, and invoices. We combine them in the form of composite patterns, which in turn make up a domain model for business-to-consumer e-commerce. We also indicate how to add security constraints to this model. This domain model can be used as a computation-independent model from which specific applications can be produced using a model-driven architecture approach.*

## 1. INTRODUCTION

Model-Driven Architecture (MDA) is an approach to application modeling and code generation [1, 2]. It starts from a Computation-Independent Model (CIM) of a system, which describes the domain and requirements. The CIM encapsulates the semantics of the problem solved by the application and this semantics should be carried along the model transformations and reflected in the code. The CIM is also useful for building applications in a more conventional way.

The use of patterns is a promising avenue to let inexperienced designers build conceptual models and as a tool for building domain models. Analysis patterns [3] are conceptual structures that capture the experience of analysts and constitute reusable elements of analysis models. Semantic Analysis Patterns (SAPs) [4] are an extension of analysis patterns to include more problem semantics. They emphasize basic functional aspects of the application model and serve as a starting point when translating requirements into a conceptual model. This type of pattern represents a minimum application (a set of basic use cases) so that it can be applied to a variety of situations and it can be combined with other related patterns to describe more complex applications.

It has been proposed to apply analysis patterns through specialization when there are abstract patterns, or through analogy when there are patterns from another domain [4]. There was also the attempt to integrate analysis patterns into MDA [5]. Hamza and Fayad have proposed stable analysis patterns as a way of developing and utilizing analysis patterns in building software systems [6], [7]. Stable analysis patterns are designed to satisfy the criteria of traceability and generality. Bobkowska and Grabowski have examined the roles of analysis patterns and analyzed their strengths and weaknesses [8]. They've also presented a case study of applying analysis patterns in system analysis.

This paper demonstrates a method of creating CIMs by using SAPs. Individual SAPs related to e-commerce are identified and combined into a domain model for business-to-consumer e-commerce applications. We also illustrate how security can be added to one of the patterns of the domain model. The domain model includes such aspects as shopping carts, products, catalogs, customer accounts, shipment, invoices and inventories. The individual patterns have been published before [9, 10, 11, 12], but corrections and extensions have been made here. The component patterns of the individual SAPs are not presented here.

The paper is structured as follows. Section 2 to 6 present the Catalog pattern, the Shopping Cart pattern, the Invoice pattern, the Order and Shipment pattern, and the Stock Manager pattern respectively. Section 7 presents a domain model for business-to-consumer e-commerce. Section 8 discusses how to add security constraints in patterns of this type. Section 9 discusses related work to end with some conclusions in Section 10.

## 2. CATALOG PATTERN

### 2.1 Intent

How can users find conveniently what products are available?

### 2.2 Context

E-commerce systems where customers can buy products

### 2.3 Problem

Web shops sell a variety of products, sometimes totally unrelated, e.g., books and food. An important problem is: How to organize product information, provide on-line guidance to the users, and improve the attraction of the web site so that users are willing to visit and return?

### 2.4 Forces

The solution is affected by the following forces:

- We need to classify and describe a variety of products so that customers can find easily what they want.
- We should be able to provide more detailed information about products of interest to a customer.
- We should be able to relate products so that we can recommend similar products to customers when they buy a product.
- We should keep customers informed of new products or changes in product prices or availability.

### 2.5 Solution

Figure 1 shows the class diagram for Catalog pattern. A **Catalog** is a collection of products. The **Product** class defines the type of product being sold, it contains the basic attributes of each product. In particular, a status attribute indicates special aspects, e.g., a new product. New products may be separated from the regular products and made known to the customers [13]. The **ProductInfo** class provides more detailed information about a product. It may also include comparison among different varieties of the same product, different brands, or provide the best price/performance ratio. Similar products are modeled as a self-association of class **Product**. Modifications to the products are notified to the customers by email to let them know there is some new or interesting product. A class **ProductObserver** watches for changes and notifies customers. Note that these classes are a type of Observer pattern. Class **Notification** keeps records of notifications sent to users. Figure 2 shows the collaborations triggered when products are modified.

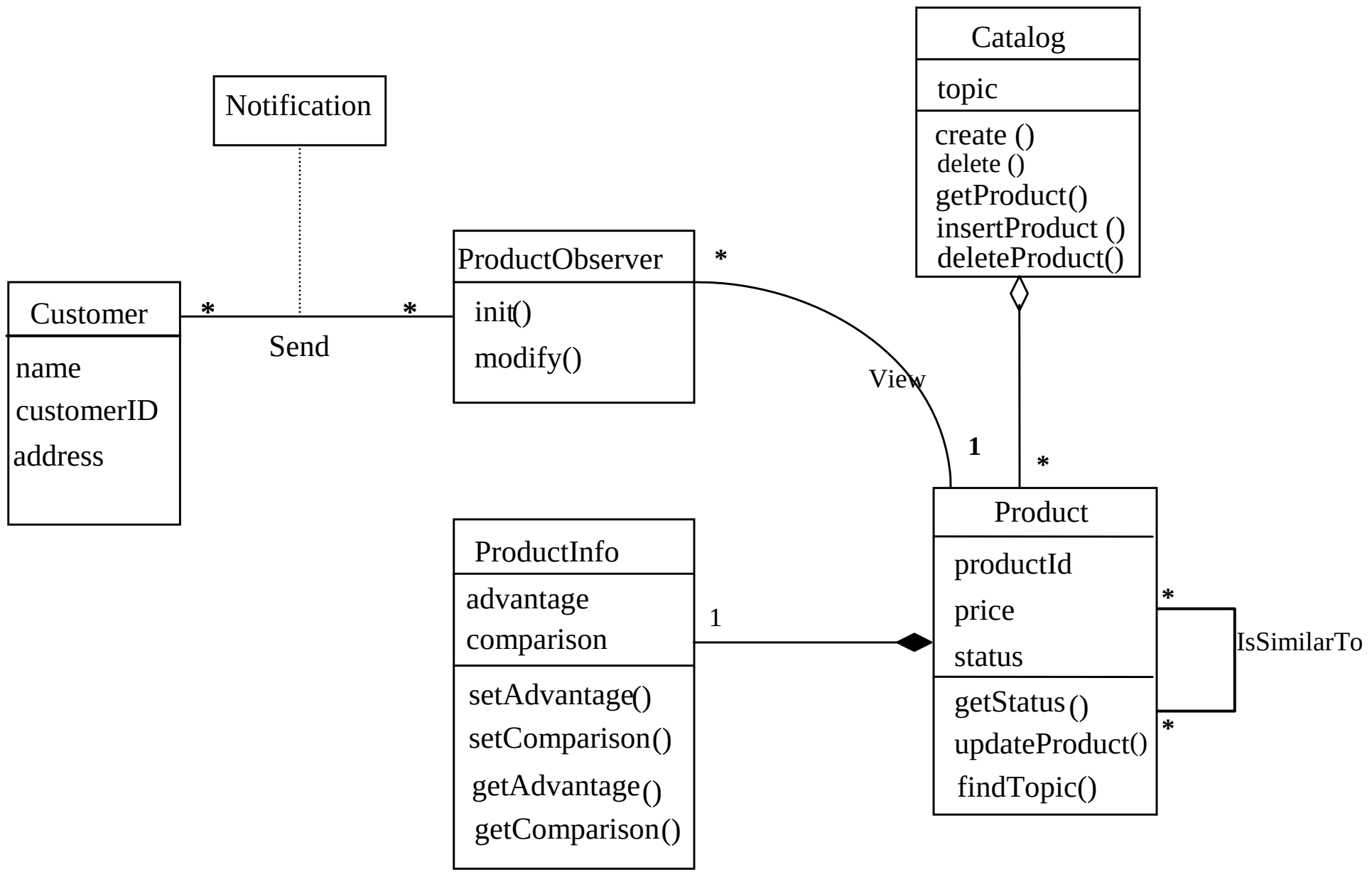

Figure 1. Class diagram for Catalog pattern

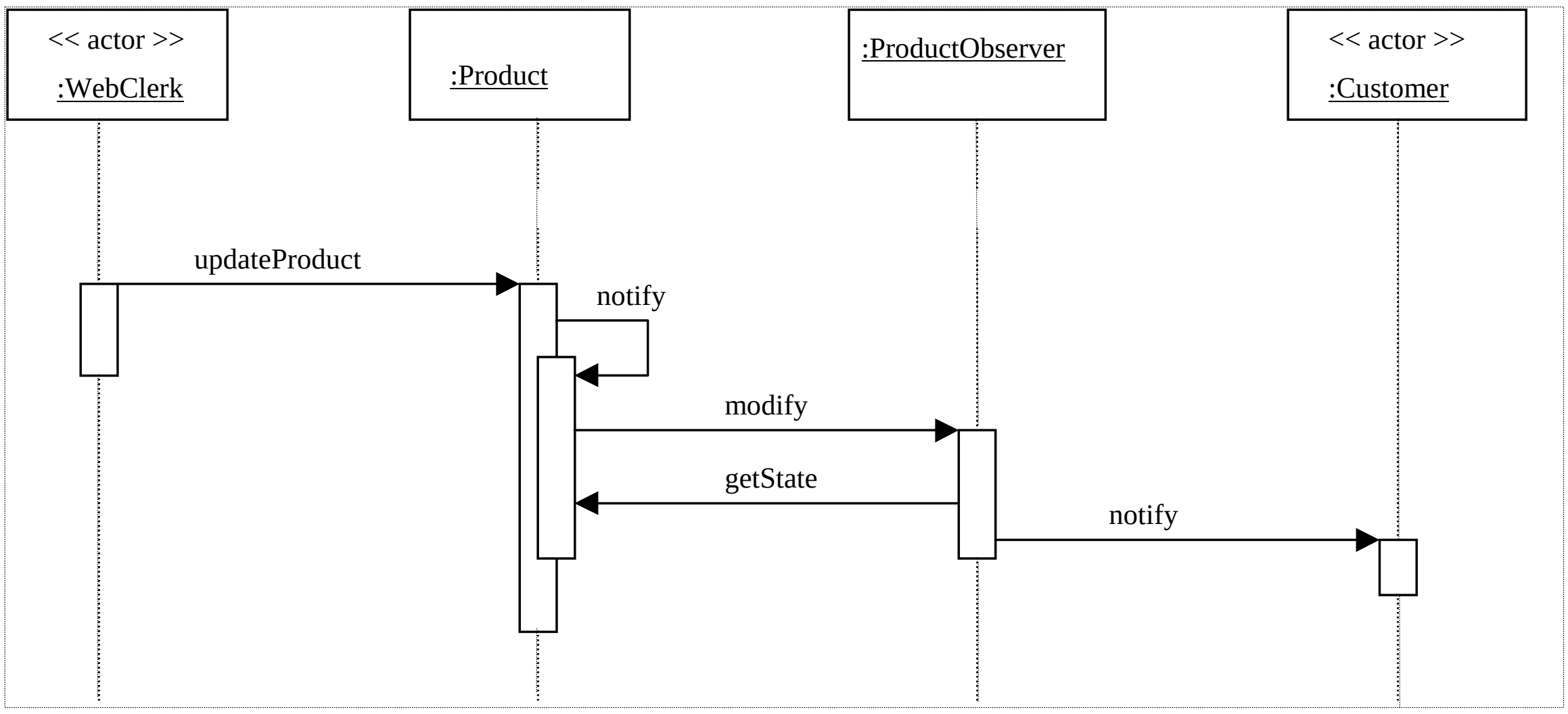

Figure 2. Sequence diagram for updating a product

## 2.6 Consequences

This pattern provides the following advantages:

- This pattern provides the needed infrastructure to describe products conveniently. Navigational classes can be added to show this information in attractive ways.

- The catalogue can classify products according to different criteria to allow convenient search and selection.
- The pattern provides a way to find more detailed information about products.
- Products can be related because they have a similar topic or because we keep statistics of products bought by people.
- We can keep customers informed of any changes to the products.
- This pattern is scalable. It is suitable for various kinds of web shops, from web malls to personal stores.
- This pattern refers to product types. Some sites sell individual products, e.g., a used-car site, an auction site. In this case the Product class becomes a set of individual products and we need an additional class to indicate the product type.

The pattern also has the following liability: It is rather complex for small shops.

### 2.7 Known uses

- On-line book stores, music stores, shoe stores, wine stores, etc.
- Most Web Application servers, e.g., IBM's WebSphere Commerce Suite, incorporate catalogs [14].
- Amazon.com

## 3. Shopping Cart Pattern

### 3.1 Intent

This pattern describes web shopping using a shopping cart

### 3.2 Context

E-Commerce systems where customers can buy products

### 3.3 Problem

Customers can select and purchase different products for a web shop. The shopping process must have well defined steps. This is necessary because we need to show the customer where he is in the process. The problem is now: How to describe the shopping process in a precise way?

### 3.4 Forces

The solution is affected by the following forces:

- We should show clearly to the customer which items she has selected and their individual and total cost.
- We should create an order and its corresponding invoice for the selected items.
- We should keep information about the buying habits of the customers so we can offer them a better service.
- We should reward loyal customers with discounts and special offers.

### 3.5 Solution

The most common metaphor for the shopping process is based on the concept of shopping cart, analogous to the carts used in supermarkets. Figure 3 shows the class diagram for the Shopping Cart pattern. The **ShoppingCart** class collects information about all the products a customer has selected. The **CartItem** object indicates the quantity and the product selected by a customer. A customer can query the products in his cart and remove products from the cart. The **Customer** class indicates the customer responsible for a shopping cart. When the shopping cart is checked out, an order and an invoice will be generated (the **Order** and **Invoice** classes). Figure 4 shows the process of selecting and checking out a product.

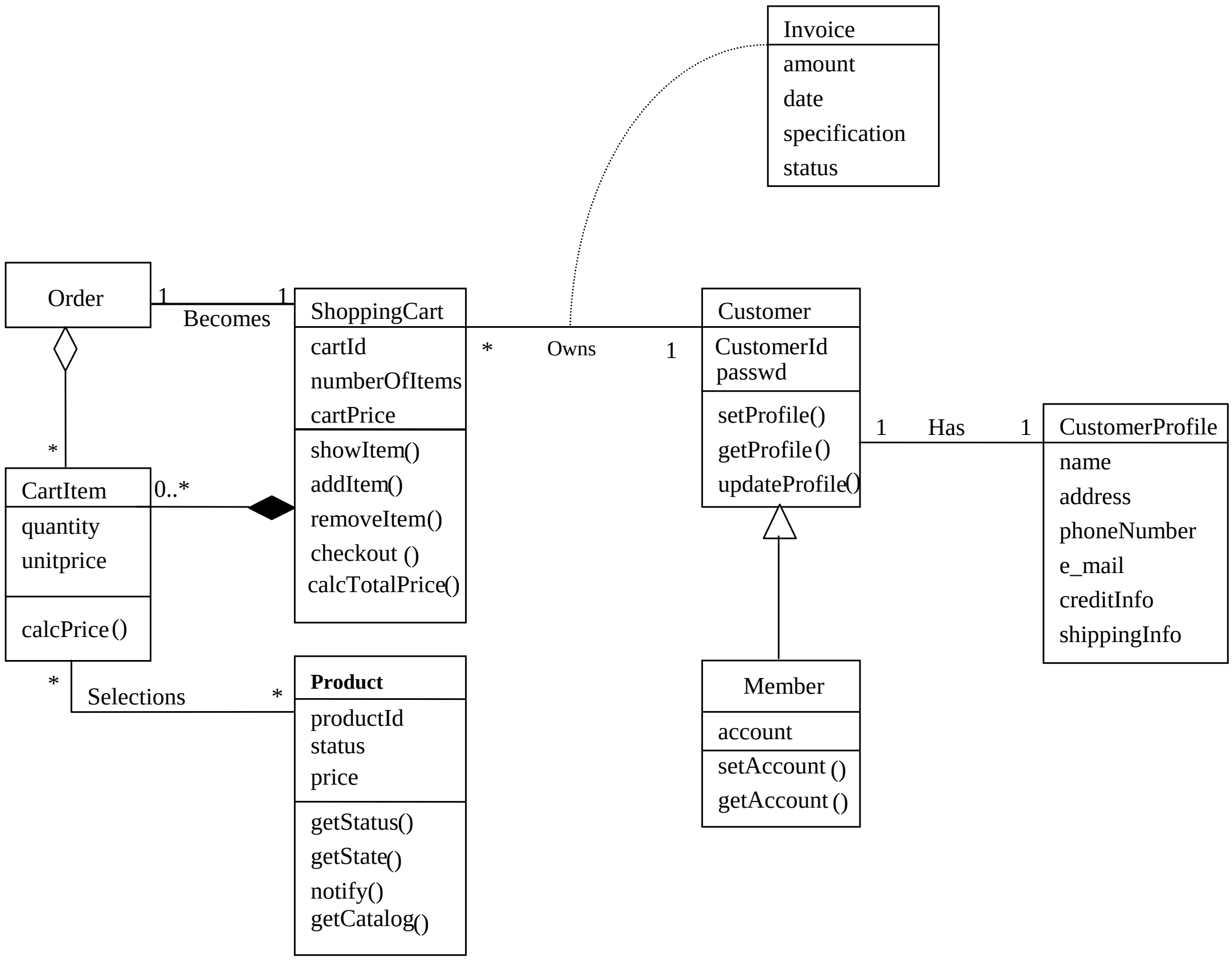

Figure 3. Class diagram for Shopping Cart pattern

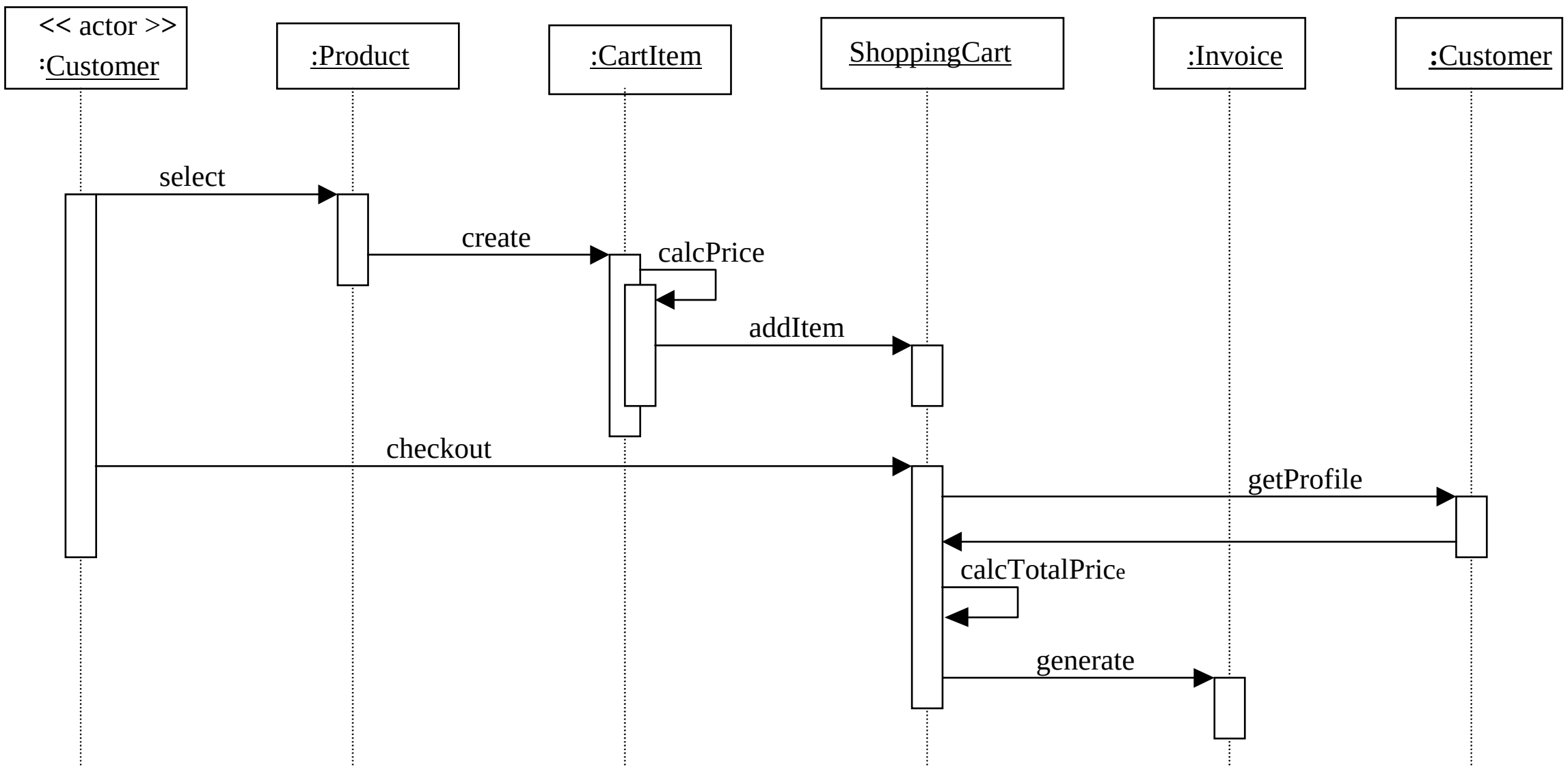

Figure 4. Sequence diagram for buying and checking out a product

### 3.6 Consequences

- This pattern describes an abstract shopping process. It provides common elements for building shops in the Internet. It can be applied to various web shops from those selling shoes to those selling software.
- The pattern should help reduce the complexity of the shopping process in the Internet. All or some of the process steps can be displayed to the user for his guidance.
- The pattern does not provide a mechanism to prevent errors, such as wrong credit card number, errors in billing address or shipping address.
- We can show what items the customer has selected and their unit and total prices.
- At the end of the process the contents of the shopping cart becomes an order to be shipped to the customer and the corresponding invoice is created.
- We can keep detailed information about the shopping habits of our customers.
- Loyal customers (members of our loyalty club) can receive discounts of coupons to buy new items.

### 3.7 Known uses

Amazon.com, Borders.com, barnesandnoble.com

## 4. Invoice Pattern

### 4.1 Intent

The pattern describes events such as the creation and validation of an invoice, followed by the payment process.

### 4.2 Context

Institutions or enterprises that require payment for products or services

## 4.3 Problem

There are many systems where we need to combine the functions of creating and preparing an invoice, and paying that invoice, including the corresponding validations. How do we represent this process in a general and abstract manner?

## 4.4 Forces

The solution is affected by the following forces:

- The creation, preparation, and validation of an invoice requires specific actors, actions in specific sequences, and must follow specific rules.
- Preparation and validation should be done by different people (separation of duty).
- There should be flexibility about who is responsible for a payment and how the payments will be made.
- The system and the client need a convenient way of keeping track of the payments made.
- Validation of every prepared invoice and every received payment has to be made to ensure that the client's information is correct and in accordance to the requirements and regulations of each system.
- We need to keep track of who created an invoice, who validated it, and who validated a payment.

## 4.5 Solution

Figure 5 is the class diagram for the Invoice pattern. Class **InvoiceCreator** defines an interface for creating an invoice. It also provides a way of preparing the invoice by adding or deleting items from it, specifying different properties, which are used to derive the final scope of the invoice. Class **Invoice** represents the document in which all the goods or services are incorporated together with the nature of each item. Class **InvoiceValidator** is used to ensure that the invoice that resulted from the steps described above is in a consistent form that complies with the trade usage. Classes **BillingPolicy** and **ValidationRule** include business policies that apply to the preparation and validation of invoices, and validation of payment.The **Payment** class represents the payment made by the client for the products and/or services incorporated in the invoice. Class **PaymentValidator** is used to validate payments according to validation rules. Class **Employee** keeps track of who validated a payment, and class **Customer** represents the customer that makes payments for the given invoice.

Figure 6 shows the sequence diagram for creating, preparing and validating an invoice. Figure 7 shows the sequences diagram for payment of an invoice. Figure 8 shows the activity diagram of creating and paying for an invoice.

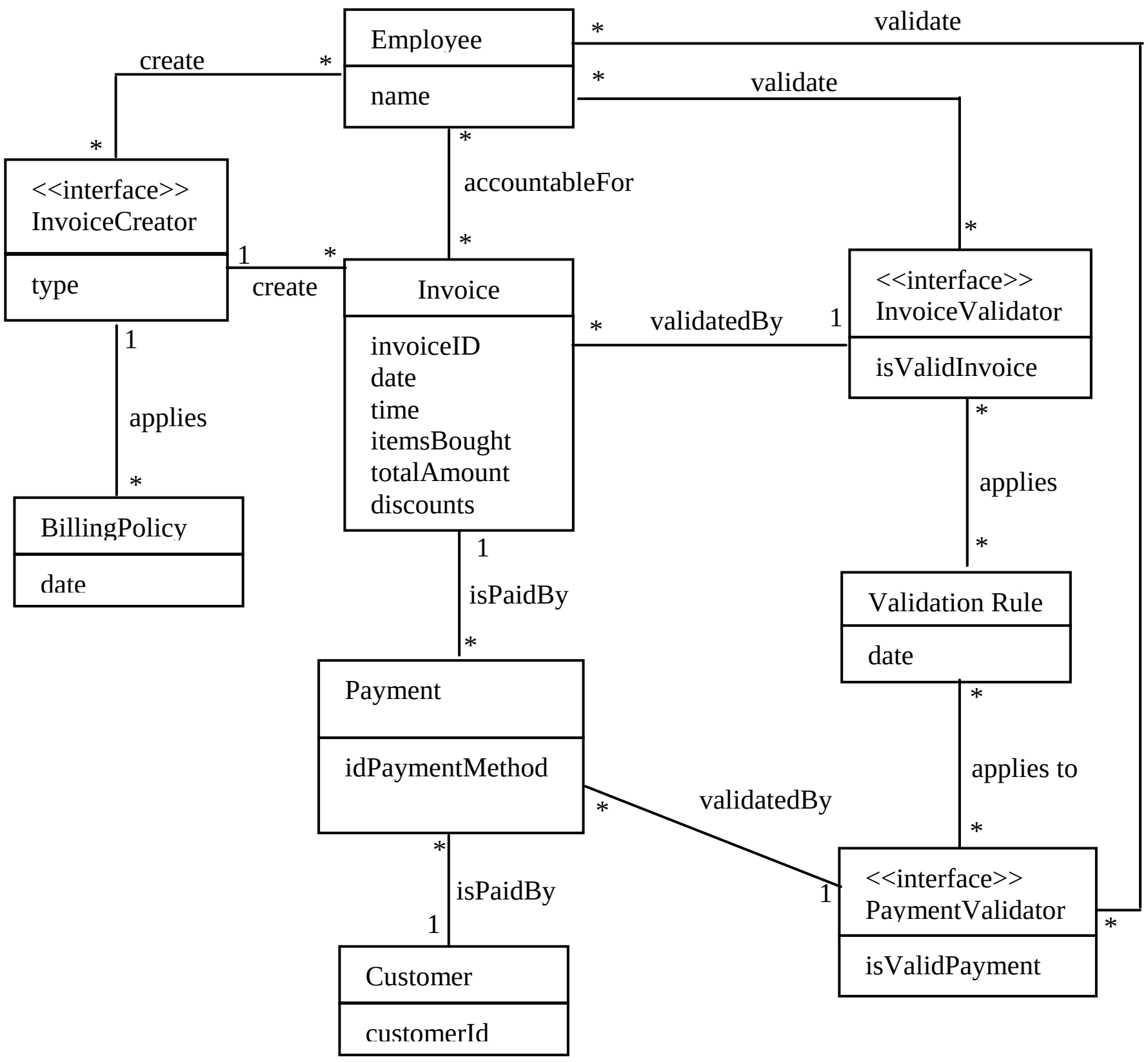

Figure 5. Class diagram for Invoice pattern

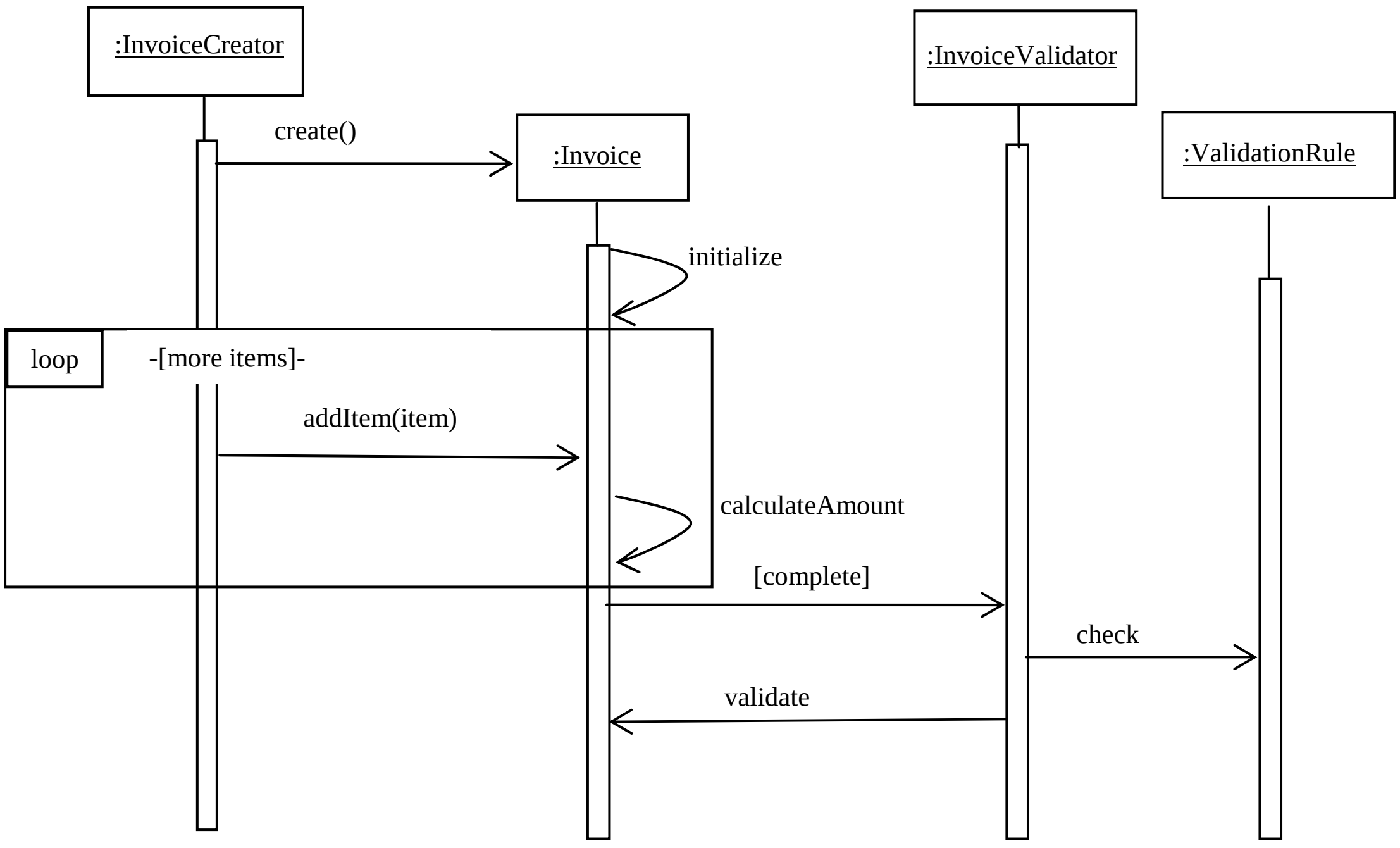

Figure 6. Sequence diagram for creating, preparing and validating an invoice

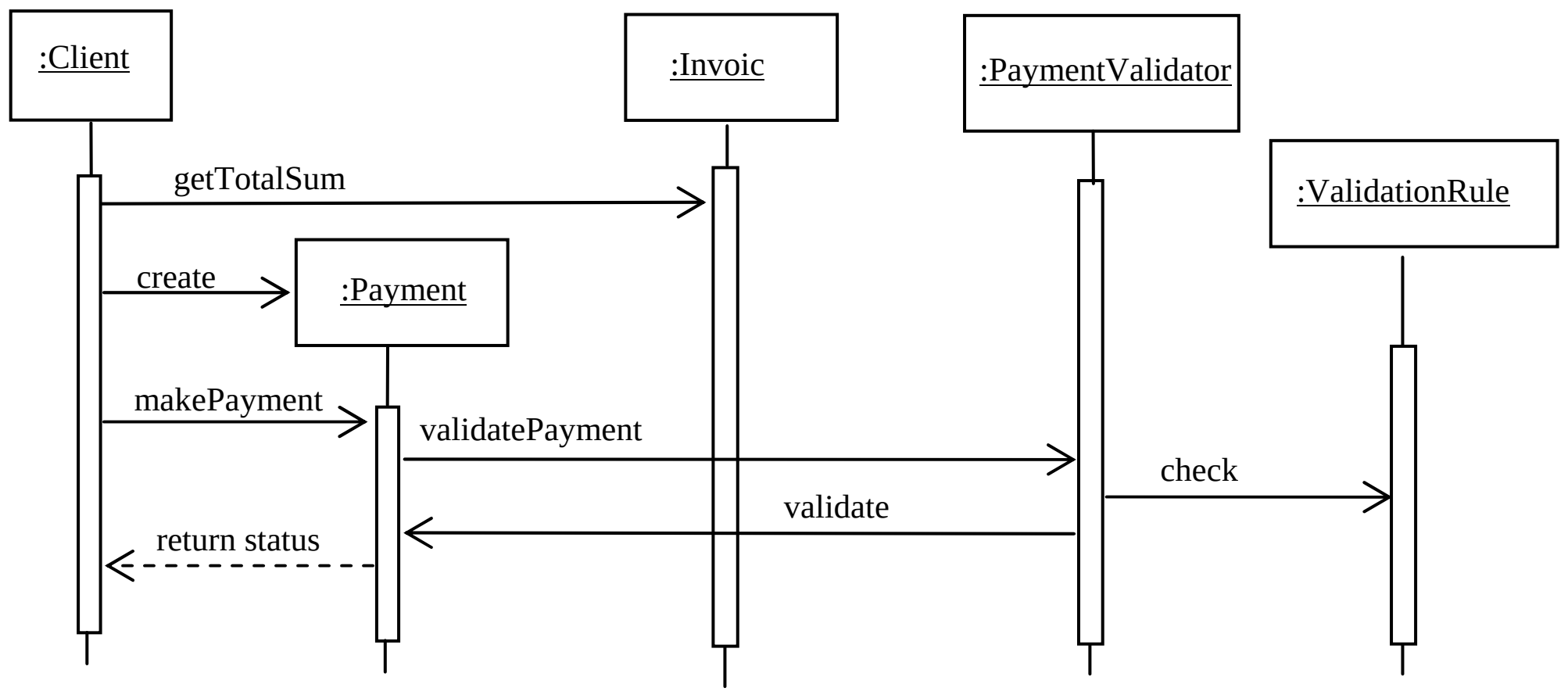

Figure 7. Sequence diagram for invoice payment

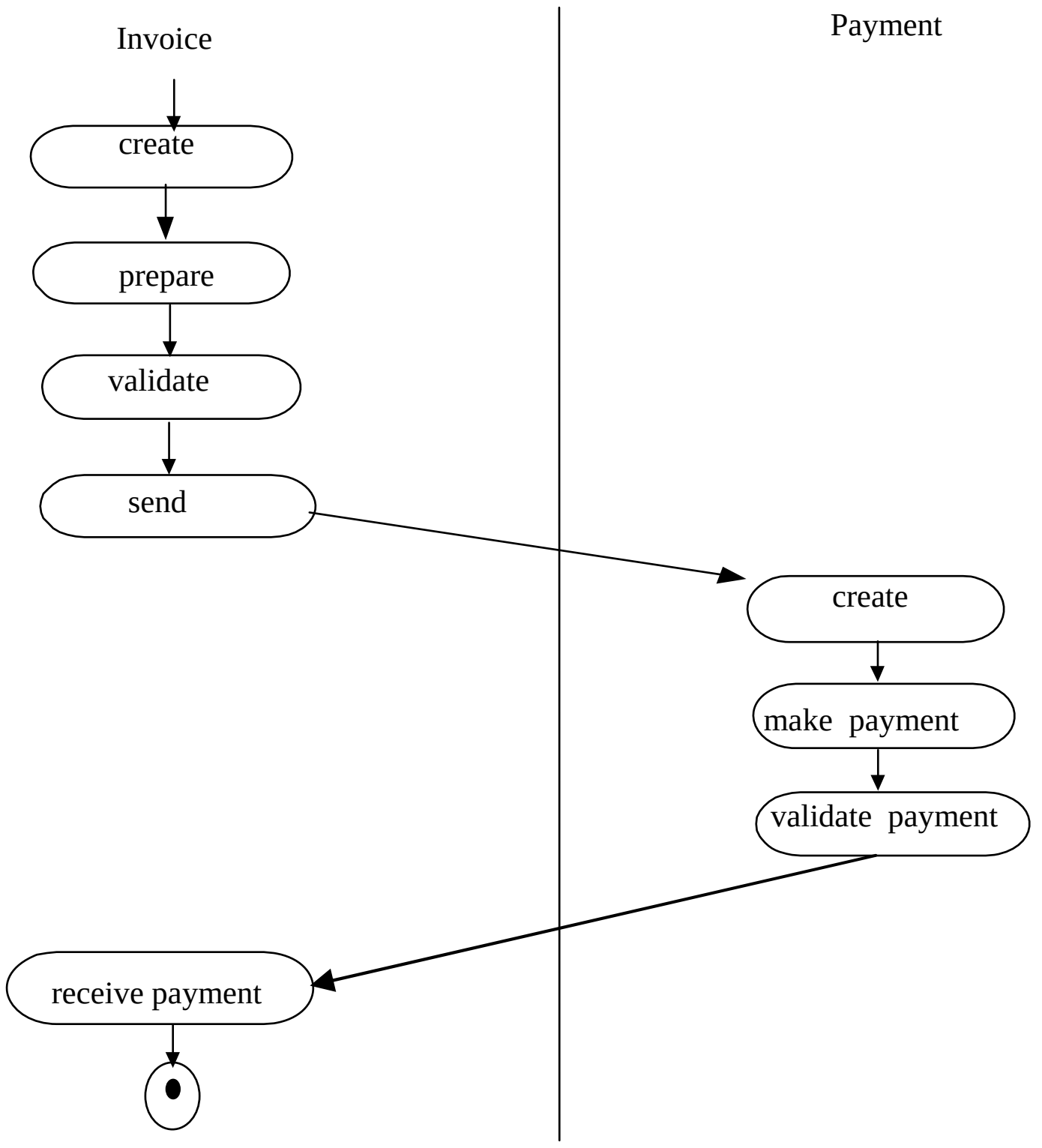

Figure 8. Activity diagram for invoice creation and payment

## 4.6 Consequences

- This pattern describes an abstract invoice preparation process that can be tailored to different specific situations.
- We can keep track of who prepared and who validated an invoice, as well as who validated payments.
- We can apply different validation rules to an invoice and a payment.

### 4.7 Known uses

- A point-of-sale system in any department store that sells products, such as Macy's.
- An on-line shopping store, where people use the Internet to log onto an on-line shop in order to buy different items, e.g. Amazon.
- Monthly invoicing for telephone or internet service, e.g. Comcast.
- SAP has an Invoice management product where they create and prepare invoices.

## 5. ORDER AND SHIPMENT PATTERN

### 5.1 Intent

This pattern describes the placement of an order for some product and its corresponding fulfillment.

### 5.2 Context

E-Commerce systems where customers can buy products

### 5.3 Problem

How to describe the process of placing an order for some product and fulfilling the order.

### 5.4 Forces

The solution is affected by the following forces:

- The institution needs to track order fulfilment to maintain customer satisfaction.
- The model must include representations of real-life documents, e.g., Orders, Line Items, and Invoices.
- Equivalent products may be substituted for requested products.

### 5.5 Solution

Figure 9 shows the class diagram for Order and Shipment pattern. The association between class **Shipment** and class **Order** shows that each shipment has a corresponding order, but an order does not necessarily result in a shipment (e.g. the order could be cancelled). The **Invoice** class describes the invoices created for each shipment. The diagram also shows that not all products ordered may be in the final shipment or that some of these products may be different from those ordered. Figure 10 and 11 show the sequence diagram and activity diagram for ordering and receiving a product.

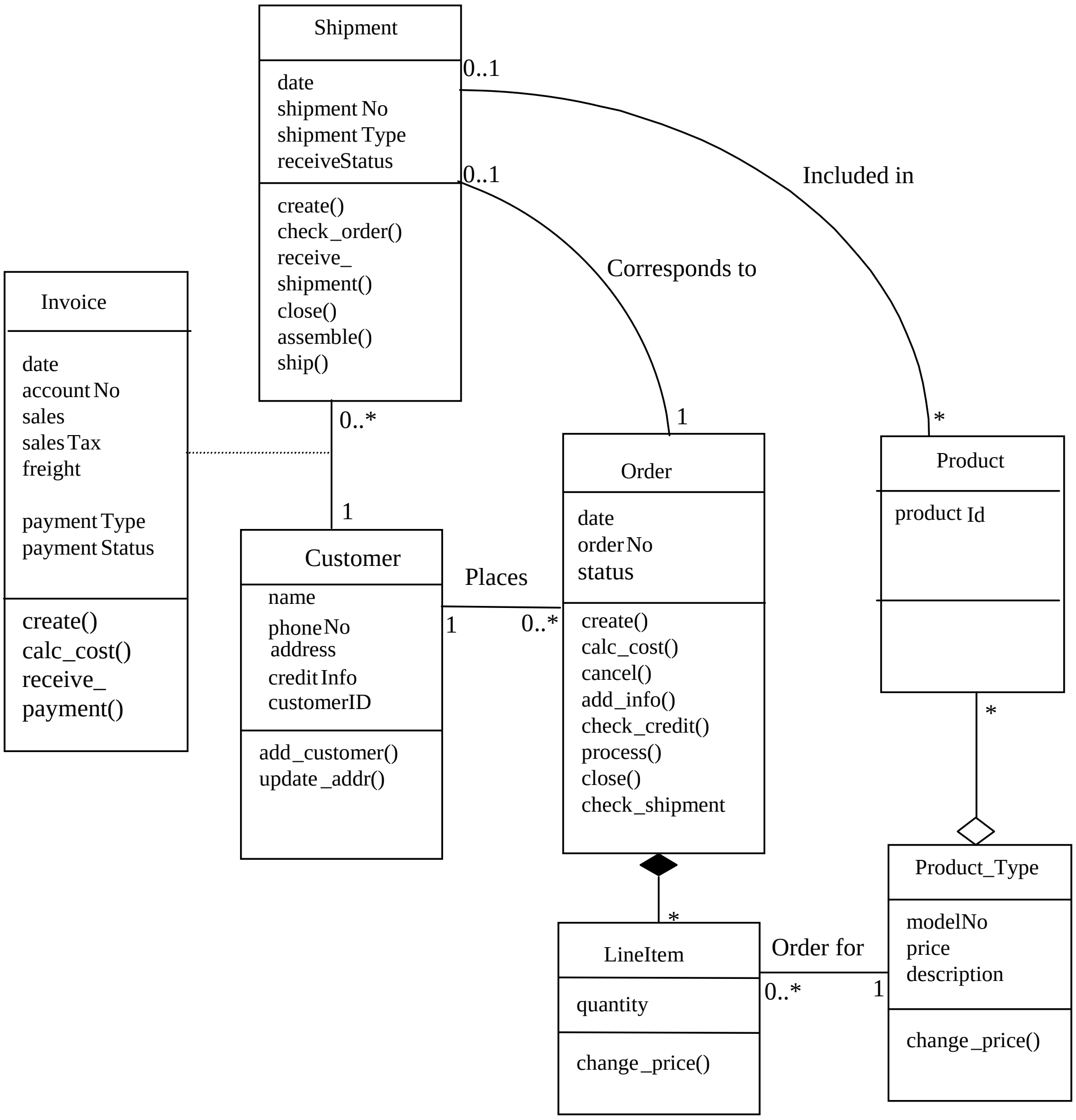

Figure 9. Class diagram for Order and Shipment pattern

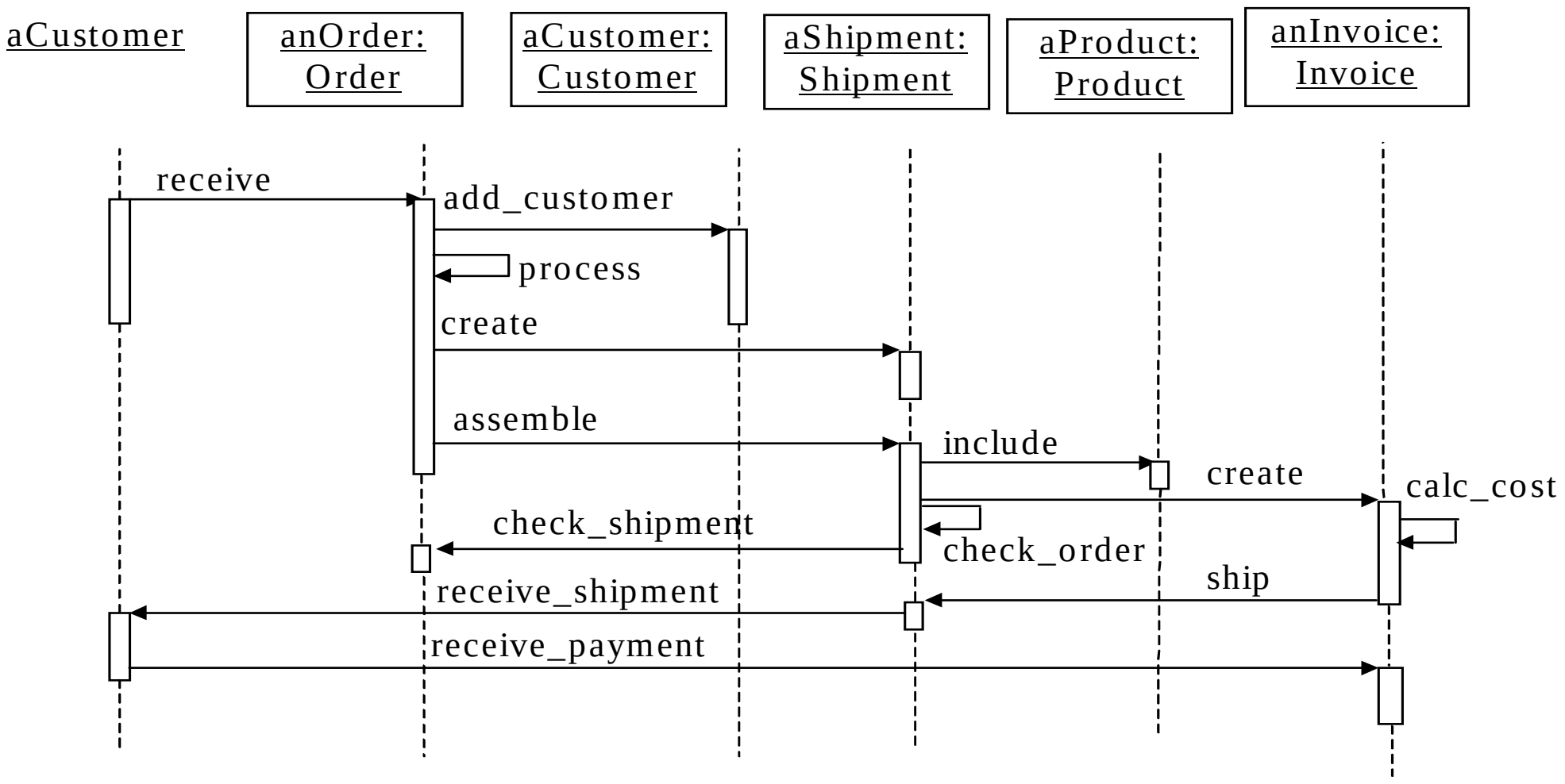

Figure 10. Sequence diagram for ordering and receiving a product

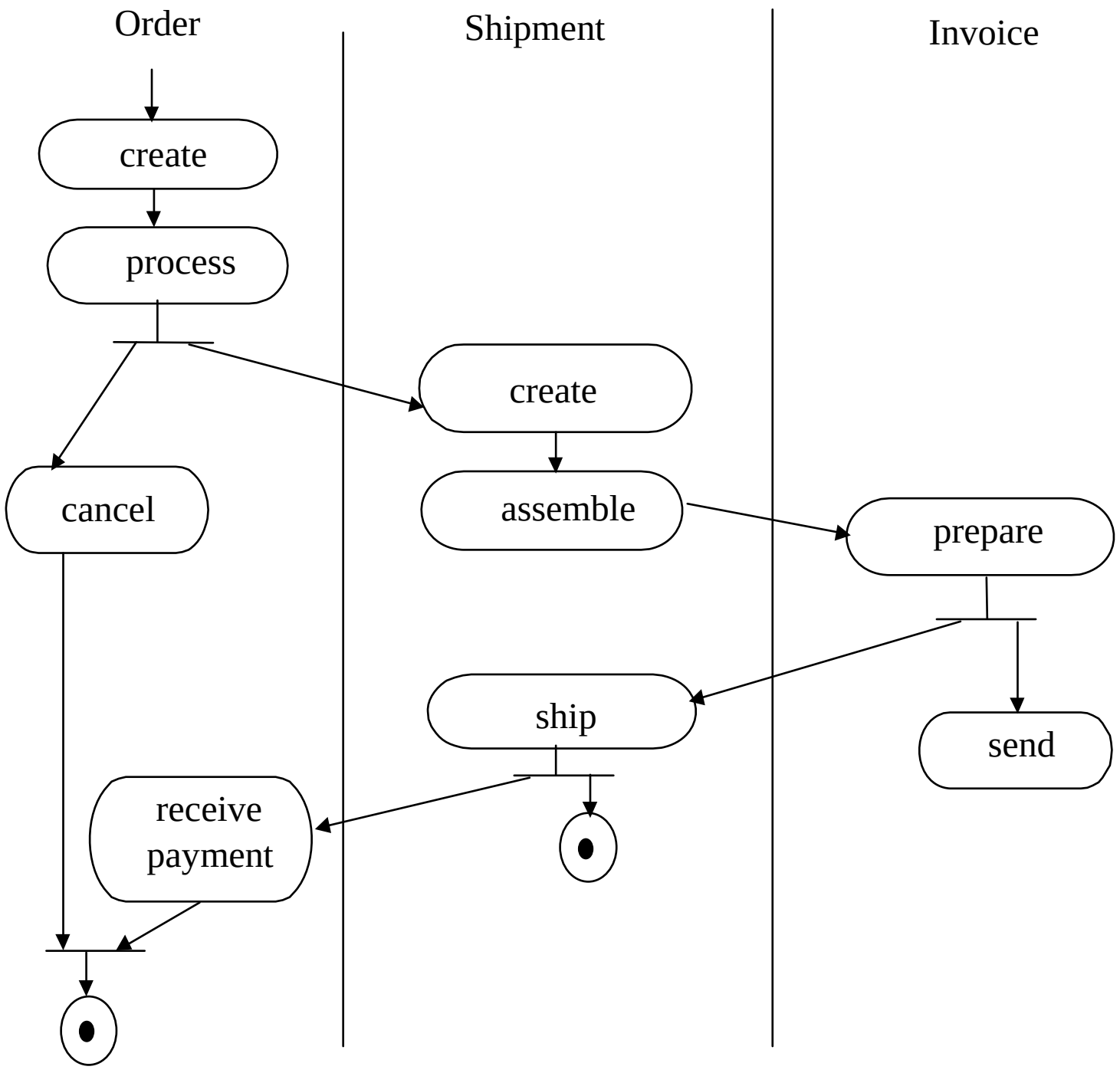

Figure 11. Activity diagram for order and shipment

### 5.6 Consequences

- An “order” is always generated in the “system” based on the customer’s expressed need.
- An order is always linked to a customer.
- Some type of documentation is always generated, with a copy archived and a copy delivered to the customer along with the product.
- The ordered product could be an individual unit, not a type.
- The ordering or receiving customers could be another subsystem or system.
- Each shipment can be related to its corresponding order.
- The receiver may or may not be the same customer who placed the order.
- The model also applies to services, with class Shipment representing the delivery of the requested service.

### 5.7 Know uses

- A retailer/service provider of some type of devices, e.g., pagers, orders a quantity of devices to be resold at retail.
- A customer orders food from a restaurant.
- A customer orders a product from an e-commerce company, e.g. Amazon.com

## 6. STOCK MANAGER PATTERN

### 6.1 Intent

This pattern keeps track of quantity and location of items in stock, and updates these quantities according to the different stages of manufacturing or production.

## 6.2 Context

Institutions or manufacturing system that use components to build products

## 6.3 Problem

How can businesses, manufacturing shops, libraries, etc., keep track of their stocks of items of different types and their locations?

## 6.4 Forces

The solution is affected by the following forces:

- In a company, items can be materials used for manufacturing or finished products. There is the possibility of losses or destruction of this stock. The institution must be able to keep track of the actual number of items in stock.
- Other functional units may change the stock quantities; i.e., any transfer or use of items anywhere should update the corresponding inventory quantities.
- The solution must describe a fundamental semantic unit. This means the solution must be simple enough to apply to a variety of situations. This is the basis for reusability.
- The solution must include representations of real-life documents.

## 6.5 Solution

Figure 12 shows the class diagram for the Stock Manager pattern. Items are classified into two varieties: finished products and components (used in product manufacturing). There exists many differences in the management of these two entities. Classes **Stock** and **Component/Product** are related by a composition association. The quantities, described by the Inventory class, are a joint property of **Stock** and **Component/Product** that has different values for different links. This model permits a designer to define different types of stock as separate collections; e.g., stock of components, stock of products. Different types of inventories can be generalized into a class **Inventory.**

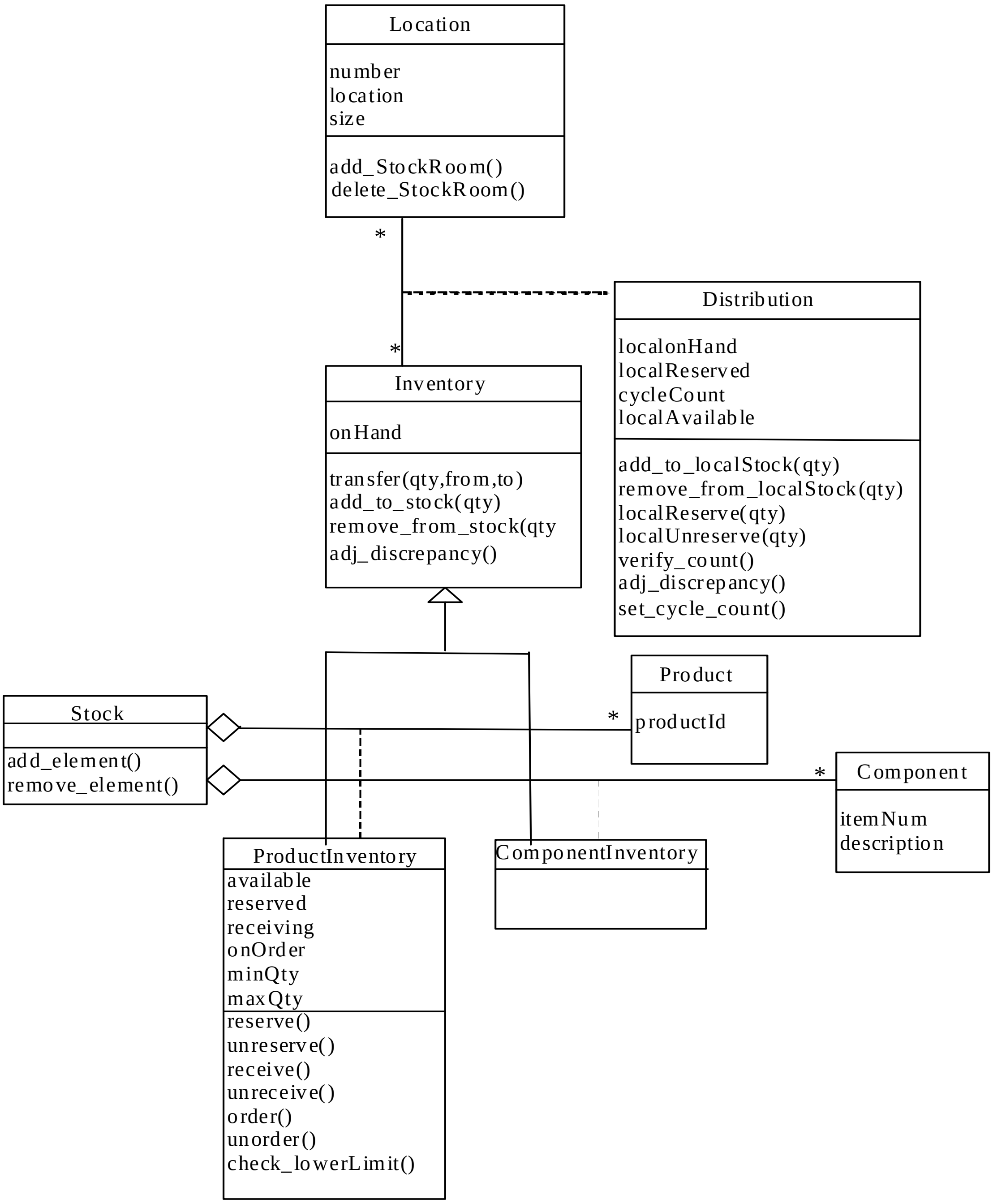

Figure 12. Class diagram for the Stock Manager pattern

Figure 13 shows the sequence diagram for moving items in or out of stock. When components or products are moved to or out of stock, their *onHand* values are increased or reduced by applying operation *add_to_stock* or *remove_from_stock*. Operation *add_to_stock* determines the distribution of items into local stockrooms based on some predetermined criteria. When components and products are moved to or out of specific stockrooms operation *add_to_localStock* or *remove_from_localStock* increases or reduces the *localonHand* quantity. Components or products can be transferred from one stockroom to another. Figure 14 shows the process of this action.

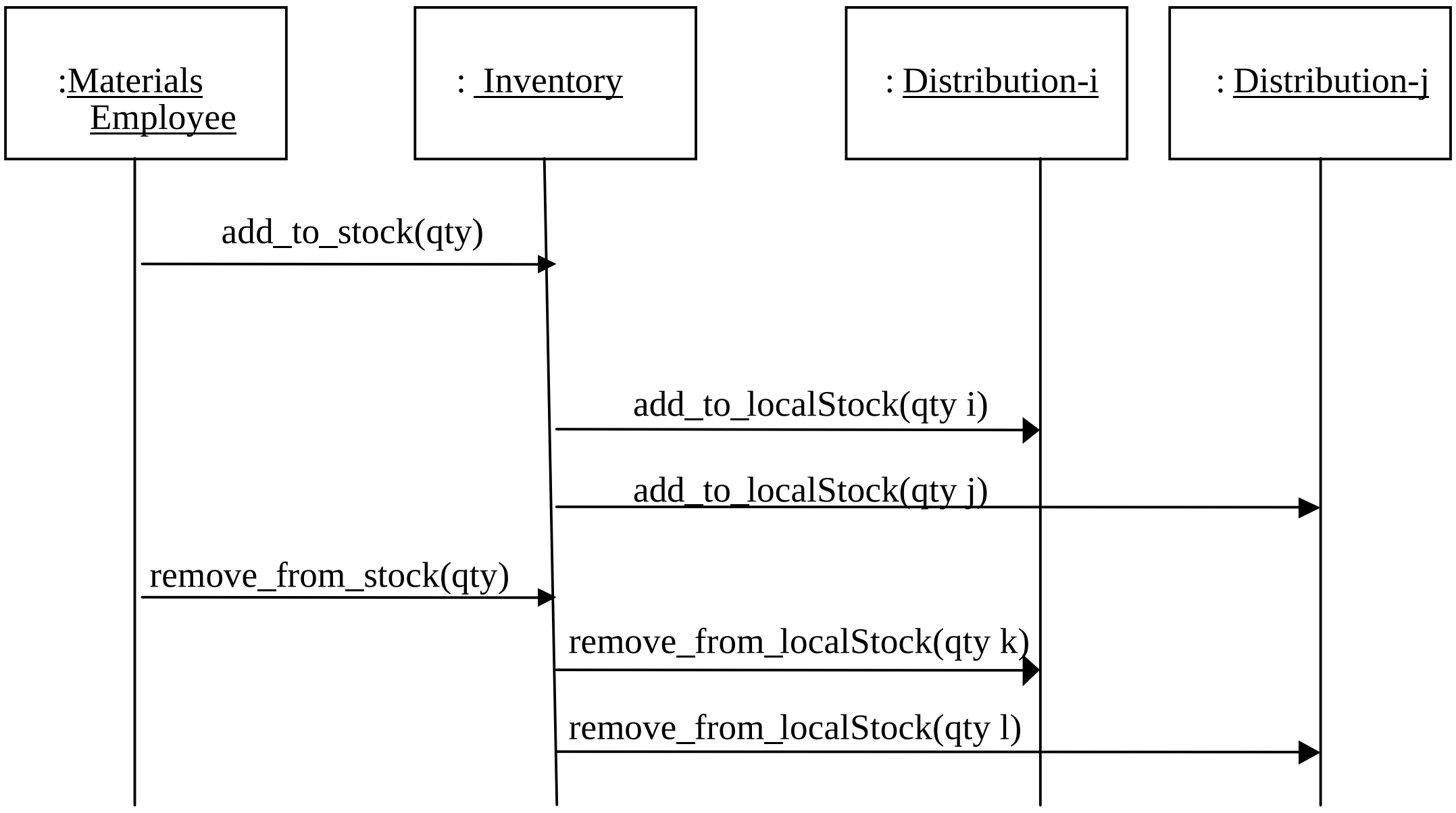

Figure 13. Sequence diagram for moving items in or out of stock

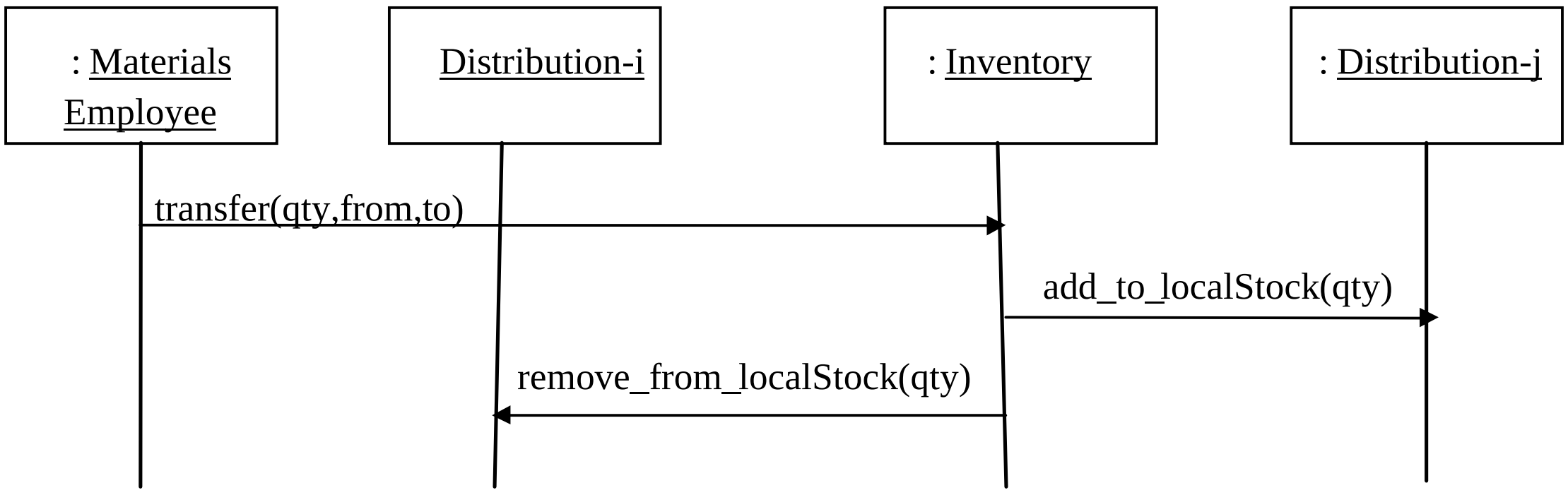

Figure 14. Sequence diagram for item transfer

Figure 15 shows the complex changes to the inventory during manufacturing. We first assume that Customer Orders have already been processed into a form that indicates in detail what and how many components are needed in the manufacturing for a type of product; this is a Shop Order. When a Materials Employee cuts the Shop Order, the values of *reserved* components are increased based on the quantity indicated by this Shop Order. When components have been physically picked from the stockroom, the values of *reserved* and *onHand* for these components are reduced. When fabrication is finished, **ProductInventory** quantities are updated by increasing their *onHand* value. Usually, a Shop Order takes several days from cut to finish. The stages cut, pick and fab let people know what is the status of the Shop Order.

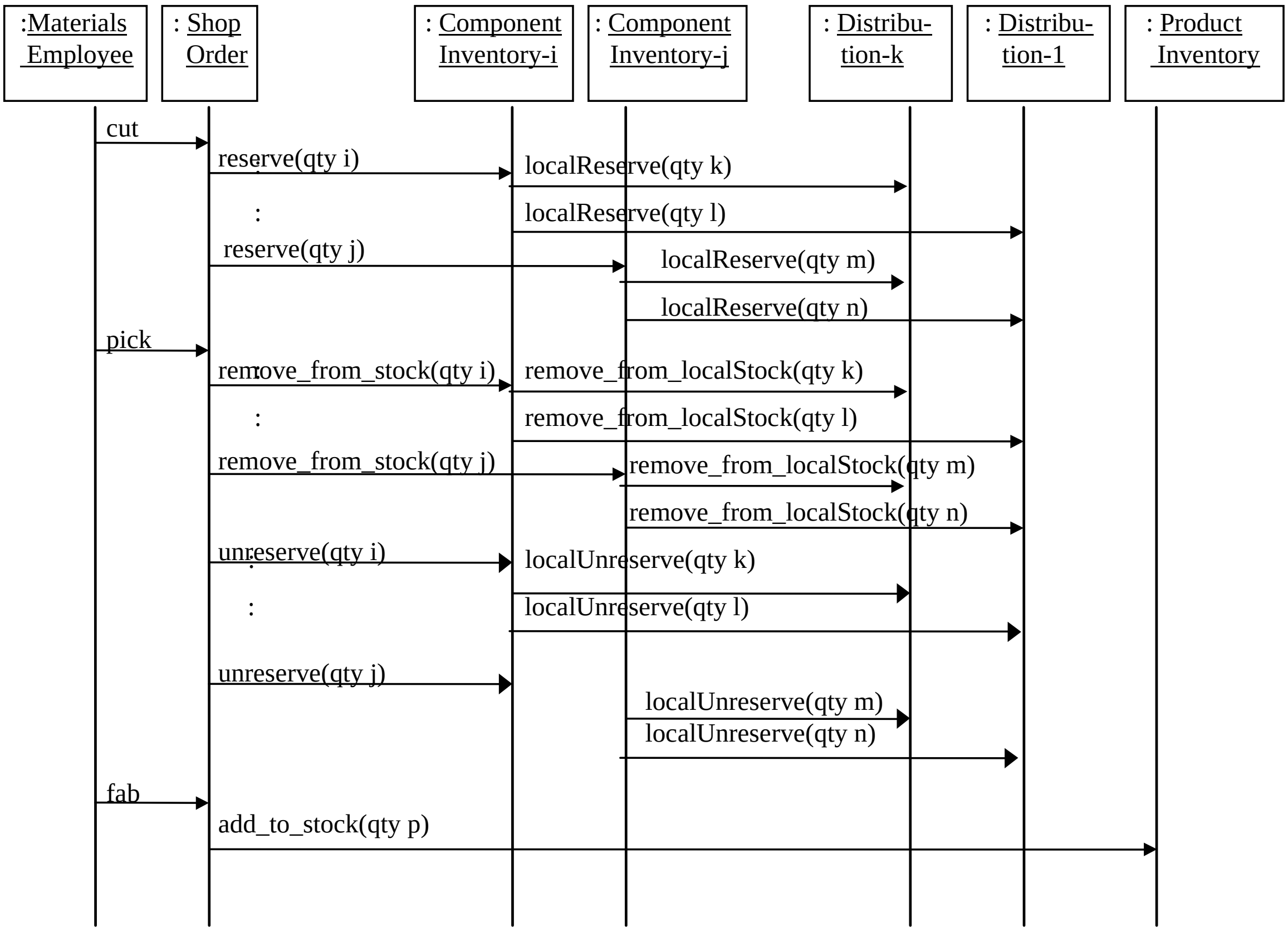

Figure 15. Sequence diagram for manufacturing a Shop Order

## 6.6 Consequences

The pattern can be used to keep track of quantities of interest for different types of stock and their distributions.

- The activities that affect the inventory may be different in different applications.
- Although the pattern is described in manufacturing terms, it can be applied to represent the inventory of a library, a business, or similar places.
- Documents such as Shop Orders and others are considered part of the external systems that interact with the inventory and are not represented in the pattern.

## 6.7 Known uses

- The model presented here was applied in the development of an inventory prototype for the Information Technology Dept. of the Pager Products Division of Motorola in Boynton Beach, FL.
- Hay's inventory model [15], includes some aspects which we left out. However, his model doesn't include dynamic aspects, attributes, or operations; it doesn't separate either stock from inventory.
- Fowler dedicates a chapter (Chapter 6 in [3]), to Inventory and Accounting.

## 7. A DOMAIN MODEL FOR E-COMMERCE

The five component patterns can be combined to develop a domain model for e-commerce applications. Each component pattern can correspond to a subsystem. Figure 16 shows how the component patterns are combined into the domain model. Classes that are in several component patterns such as Customer, Invoice are only included in one subsystem. Subsystems dependencies are also shown in the diagram. The CartItem in Shopping Cart subsystem corresponds to the LineItem in Order and Shipment subsystem, so LineItem is removed from the Order and Shipment subsystem.

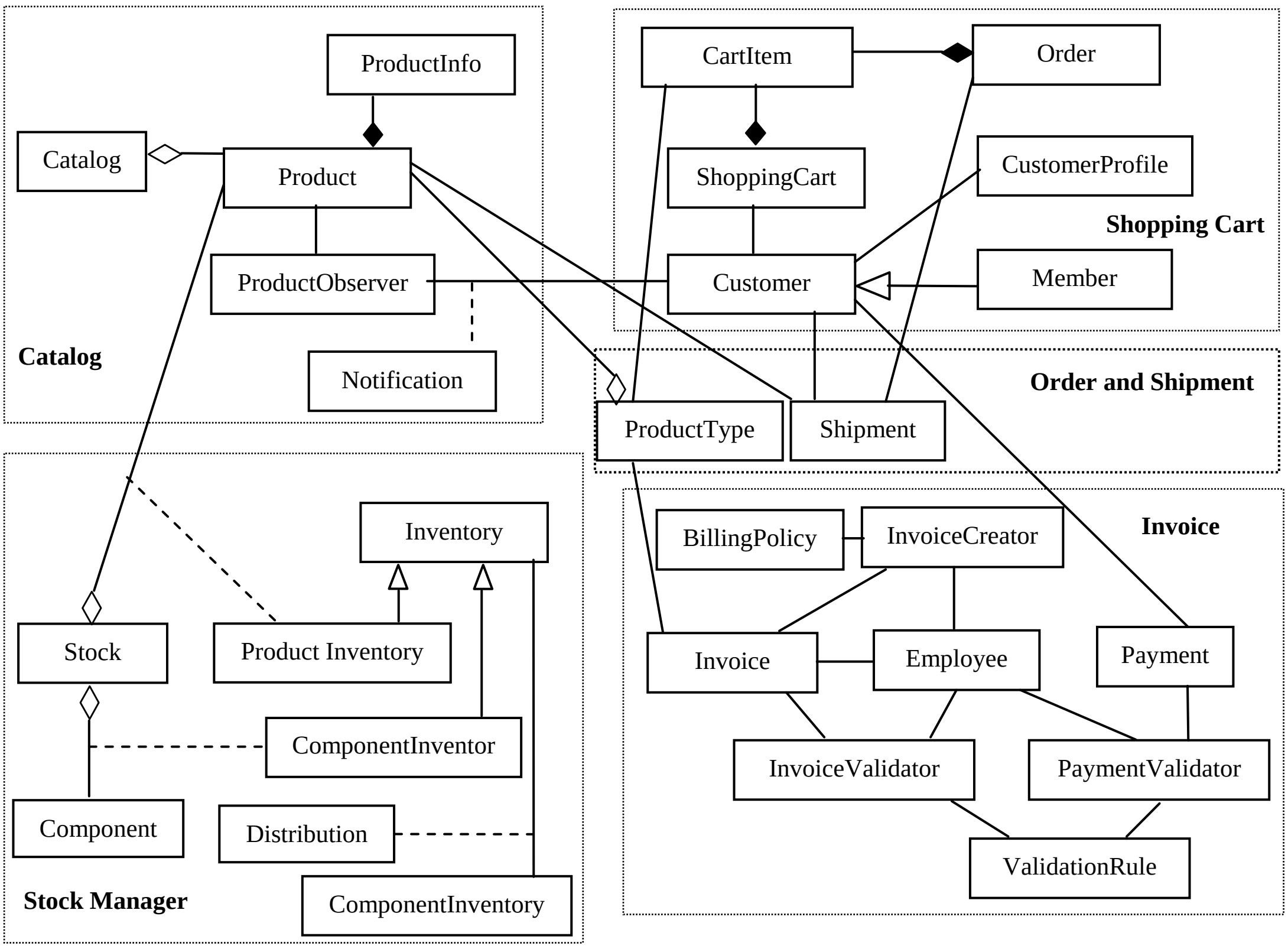

Figure 16. The domain model for e-commerce

## 8. SECURE E-COMMERCE

Security constraints can be added to each of the component patterns to produce a domain model for secure e-commerce. We demonstrate here how to add security constraints by instantiating a security pattern, that is, Role-Based Access Control (RBAC) pattern [16]. In the RBAC pattern, users are assigned to the roles according to their tasks or jobs and rights are assigned to the roles. In this way, a need-to-know policy can be applied, where roles get only the rights they need to perform their tasks. Figure 17 shows how to add security constraints to the Shopping Cart pattern by applying six instances of the RBAC pattern.

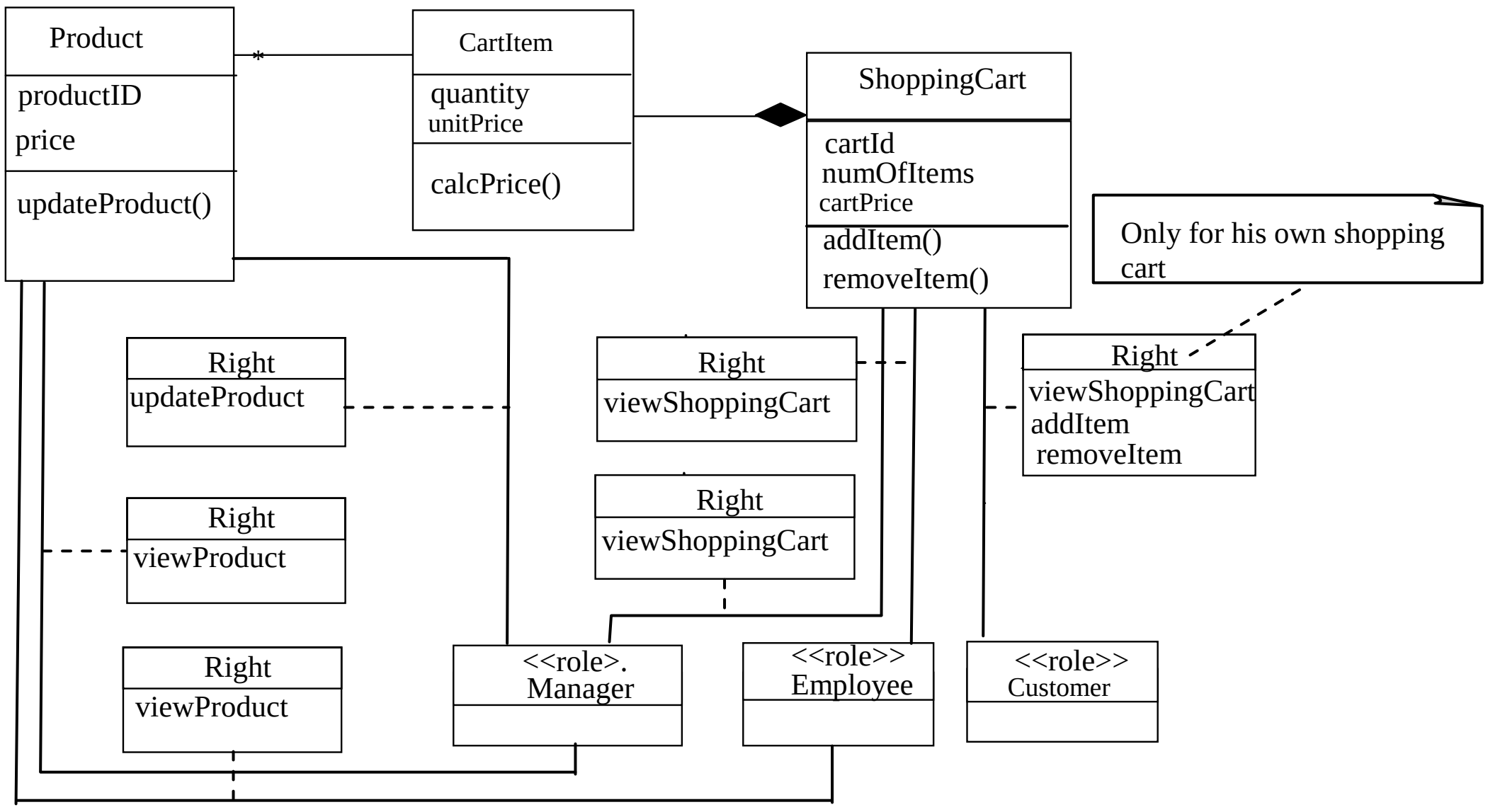

Figure 17. Adding security constraint to Shopping Cart pattern

## 9. RELATED WORK

Jazayeri and Podnar [17] presented a business and domain model for information commerce, which describes a virtual environment in which buyers and sellers of information may trade information products. The domain model was developed from scratch from use cases. The Customer, Product and Order classes in the domain model are also included in our domain model. The roles of Intermediary and Provider, as well as Request, Offer, Document are not included in our model.

A phase-structured model for e-commerce business models is presented in [18]. In this model, business processes are broken down into five phases: advertising, negotiation, ordering, payment, and delivery. A 3-party model (customer, intermediary, provider) is used to model interactions in e-commerce business models. The phase model is used to analyze the specific security requirements of e-commerce business models, highlight potential threat scenarios and describe their solutions. Our domain model covers advertising (in the Catalog patter), ordering, payment and delivery phases. Our model does not include the intermediary party.

Object-Oriented Hypermedia Design Method (OOHDM) [19] is a web application design method that includes conceptual design, navigational design, abstract interface design and implementation. In [20], OOHDM was extended to allow developers clearly specify and design web applications that embody business processes. The extended OOHDM was used to model an online retail store. Our domain model includes similar objects or cover similar aspects in their conceptual schema of the online retail store, for example, ShoppingCart, Order, Customer, Item, CD, ShippingAddress, PaymentOptions, DeliveryOptions.

Lau [21] conducted domain analysis of e-commerce systems, and applied feature-based model template development process to develop and model a business-to-consumer e-commerce system. The feature model consists of over 200 features which were divided into two categories: store front and business management. The model template includes two class diagrams and seventeen activity diagrams. Our model and their model both have objects such as Catalog, Customer, Product, ShoppingCart, CartItem, Order, Payment, etc. Their model has features that we do not have, such as wishlist, customer registration, etc. Our model includes inventory and stock management, invoice creation and validation which are not covered by their model.

## 10. CONCLUSION AND DISCUSSION

This paper presents a set of patterns appropriate for business-to-consumer e-commerce. These patterns are then combined into a domain model. This paper also shows how security constraints can be added to the domain model. The domain model can be used as a CIM from which specific applications (for example, an inventory management system) can be produced using an MDA approach. A correct CIM is fundamental for any MDA process because it defines the semantics of the problem represented by the application solution. If the details of one the component SAPs are not of interest we can replace the SAP by a Façade [22].

When we add security constraints in the CIM we can enforce that all the applications derived from it follow the same constraints along the MDA transformations [23]. Similarly, institution policies and legal regulations can be defined in the domain model. Varieties of the model can be produced for different environments requiring different regulations, for example, HIPAA regulations can be used for medical applications.

Our domain model includes the basic aspects of e-commerce. It could be extended by adding an Intermediary role, a Wishlist, etc. Other aspects that can be added include personalization [24, 25] and usability-oriented patterns [13, 26, 27].